\documentclass{article}

\usepackage{float}

\usepackage{arxiv}
\usepackage{xcolor}

\usepackage[utf8]{inputenc} % allow utf-8 input
\usepackage[T1]{fontenc}    % use 8-bit T1 fonts
\usepackage{hyperref}       % hyperlinks
\usepackage{url}            % simple URL typesetting
\usepackage{booktabs}       % professional-quality tables
\usepackage{amsfonts}       % blackboard math symbols
\usepackage{nicefrac}       % compact symbols for 1/2, etc.
\usepackage{microtype}      % microtypography
\usepackage{lipsum}
\usepackage{graphicx}
\usepackage{amsmath}
\graphicspath{ {./images/} }
\usepackage[english]{babel}

\usepackage{stackengine}

\title{Efficient Method for accelerating line searches using a combined Schur complement domain decomposition and Born series expansions in photonic-based adjoint optimization}

\author{
 Nathan Zhao \\
  Department of Applied Physics\\
  Stanford University\\
  Stanford, CA 94305 \\
  \texttt{nzz2102@stanford.edu} \\
   \And
 Salim Boutami\\
  Department of Electrical Engineering\\
  Stanford University\\
  Stanford, CA 94305 \\
  \texttt{sboutami@gmail.com} \\
  \And
 Shanhui Fan \\
  Department of Electrical Engineering\\
  Stanford University\\
  Stanford, CA 94305 \\
  \texttt{shanhui@stanford.edu} \\
}

\begin{document}
\maketitle
\begin{abstract}
    
A line search in gradient-based optimization algorithm solves the problem of determining the optimal learning rate for a given gradient or search direction in a single iteration. For most problems, this is determined by evaluating different candidate learning rates to find the optimum, which can be expensive. Recent work has provided an efficient way to perform a line search with the use of the Shanks transformation of a Born series derived from the Lippman-Schwinger formalism. In this paper we show that the cost for performing such a line search can be further reduced with the use of the method of the Schur complement domain decomposition, which can lead to a 10-fold total speed-up resulting from the reduced number of iterations to convergence and reduced wall-clock time per iteration.

\end{abstract}

% keywords can be removed

\section{Introduction}

Optimization approach based on gradient descent has been well established for the inverse design and optimization of both linear and nonlinear photonic devices \cite{Piggott2015, Liu2011, Lalau-Keraly2013, Veronis2004, Georgieva2002, Rodriguez2018, Lu2012, Zhang:21}. In each step of the optimization, one starts with a candidate structure and computes the gradient in the design parameter space of its objective function. One then generates a new candidate structure by updating all these parameters along the direction of the gradient. This process iterates until a structure with sufficiently good performance is found. What enables the efficient implementation of the gradient descent optimization is the development of adjoint variable method \cite{Lalau-Keraly2013,Plessix2006}, which allows one to determine the gradient of the objective function with respect to all design parameters with only two simulations of a given structure. 

In each step of the gradient descent optimization, in addition to the information of the gradient, one would also need to provide the distance over which the design parameters need to be varied along the gradient direction. This distance is usually controlled by a hyperparameter known as the learning rate. The optimal learning rate can in principle be determined if one performs an exact line search along the gradient direction. Any line search, however, can be demanding computationally, as it requires simulations of many structures along the gradient direction. Therefore, in practice, the learning rate is often obtained heuristically.  The use of an approximate learning rate as opposed to the optimal rate can significantly increase the number of iterations required in the gradient-based optimization in order to reach an optimal structure. For further improvement of the gradient descent optimization, it is certainly of interest to develop algorithms that can efficiently determine the optimal learning rate \cite{Boutami:20, boutami_zhao}. Moreover, given that in such optimizations many structures need to be evaluated, it would also be of interest to reduce the computational costs associated with simulating individual structures. 

Our paper builds upon two recent advances in the development of gradient descent approaches for the optimization of photonic devices. First, in Refs. \cite{Boutami:20, boutami_zhao} it was shown that the Lippman-Schwinger equation and its corresponding Born series can provide an effective way to determine the optimal learning rate, by enabling an efficient line search along the gradient direction. Second, in Ref. \cite{Zhao:19} it was shown that a Schur’s complement approach can be efficient for reducing the computational cost associated with evaluating candidate structures. In many photonics optimization problems, one only optimizes a subset of degrees of freedom inside the entire computational domain. The Schur’s complement approach can achieve significant acceleration in the evaluation of the candidate structures by reducing the solution of the problem to only those degrees of freedom being optimized.  In this paper, we show that these two advances can be combined together to significantly improve the gradient descent optimization of nanophotonic devices. 

The remainder of the paper is organized as follows. In Section 2, we provide a problem set-up and the mathematical formalism. In Section 3, we verify our results numerically using an example system where applying the line search in Ref. \cite{Boutami:20} alone may not yield any speedup. In Section 4, we analyze the nonzero fill-in of the matrices used in Section 3 to see how our line search algorithm can experience significant acceleration. We conclude in Section 5.

\section{Mathematical Formalism}
% As a brief overview of our formalism section, we  frame gradient-based optimization as three steps, a) evaluating an objective function/simulating a structure, b) computing a gradient and c) performing a parameter update with a learning rate. The mechanics of all three steps are important to understanding our formalism of accelerating gradient-based photonic optimization.
% \begin{figure}[H]
%     \centering
%     \includegraphics[width = 5 in]{figures/gradient_time.png}
%     \caption{a) the three steps required for one iteration of ordinary gradient descent. They consist of evaluating the objective function, computing the gradient, and updating the parameters. The size of each block are proportional to the amount of time it takes for each step. b) the four steps required for gradient descent plus a line search (green), which requires evaluating the objective function again multiple times.}
%     \label{fig:linesearch}
% \end{figure}

\subsection{Evaluating the Objective Function: Solving Maxwell's equations}

In a typical optimization procedure of a nanophotonic device, in order to evaluate the objective function, we simulate a device by solving the Maxwell's equations, here written in the frequency domain:
\begin{equation}
\bigg(\frac{1}{\mu_0}\nabla \times \nabla \times  - \omega^2 \epsilon_0 \epsilon_r(\mathbf{r}) \bigg)\mathbf{E(r)} = -i\omega \mathbf{J(r)}
\label{eq:maxwell}
\end{equation}
In Eq. \eqref{eq:maxwell}, $\mathbf{E(r)}$ denotes the electric field, consisting of three components $E_x(\mathbf{r})$, $E_y(\mathbf{r})$, $E_z(\mathbf{r})$, $\mathbf{J(r)}$ is the source, $\epsilon_r(\mathbf{r})$ is the relative dielectric permittivity, $\epsilon_0$ is the permittivity of free space, and the magnetic permeability is assumed to be the permeability of free space $\mu_0$ in the entire computational domain. In Eq. \eqref{eq:maxwell}, $\mathbf{E(r)}$ and $\mathbf{J(r)}$ are complex functions of the spatial coordinate $\mathbf{r}$.

By discretizing Eq. \eqref{eq:maxwell} via finite differences on a Yee grid \cite{yee1966}, we can convert this into a system of linear equations:

% \begin{equation}
%       A^\top \mathbf{E} = -i\omega \mathbf{J} 
%     \label{eq:forward}
% \end{equation}
% \textcolor{red}{I'm wondering if a change of notation is merited. Usually matrices are in bold upper case font and vectors in bold lower case font. I'm wondering if we should adjust this (but at the same time, somehow the bold capital E(r) seems like it's okay. Mainly, I don't like the matrices being just plain font and scalars like L the objective function also being plain font}
\begin{equation}
      \hat A \mathbf{E} = -i\omega \mathbf{J}
    \label{eq:forward}
\end{equation}
with the solution:
\begin{equation}
    \mathbf{E} = -i\omega \hat G\mathbf{J} 
    \label{eq:solution}
\end{equation}
where we now, for simplicity, suppress the spatial coordinate vector $\mathbf{r}$. $\hat A$ is a matrix which represents the discretized approximation of the left hand side of Eq. \eqref{eq:maxwell}. ${\hat G}$ is the Green's function corresponding to $\hat A$. The $\mathbf{E}$ in Eq. \eqref{eq:forward} is now a vector over the field components and over each spatial grid point. Likewise, $\mathbf{J}$ is also a vector over all spatial grid points containing information about the sources on this discretized domain.

To illustrate the computational technique developed, in this paper we will consider a design problem shown in Fig. 2. The device includes one input and two output waveguides. All these waveguides are single-moded. These waveguides are connected by a design region. The objective of the design to achieve high-efficiency, equal splitting of the input to the two output waveguides. In the simulation, the device is surrounded by a region consisting of perfectly matched layers (PML)\cite{Berenger1994, Wonseok2012}. The modes are excited by a source in the waveguide outside the design region, and the fluxes are computed along lines perpendicular to the output waveguides to determine the transmitted flux.  For this problem, the matrix $\hat A$ describes the entire structure including the PML. $\mathbf{J}$ describes the excitation source in the input waveguide. $\mathbf{E}$ is the electric field distribution inside the computational domain. 

% As an example, consider the setup shown in Fig. \ref{fig:splitter_intro}a), which consists of a mode splitter. An input waveguide arm comes from the bottom edge and enters the design region, which has the task of splitting the power of the mode equally into the two arms in the center right and top middle of Fig.\ref{fig:splitter_intro}a). The sky blue region represents a perfectly matched layer \cite{Berenger1994}.  In Fig. \ref{fig:splitter_intro}b), we simplify the computational domain into the design region, as denoted by subscript $O$ since this is the region where the optimization takes place, and the background region, denoted by the subscript $B$, that contains the rest of the computational domain.  The input source is in the background region.
\begin{figure}[H]
    \centering
    \includegraphics[width = 4.5 in]{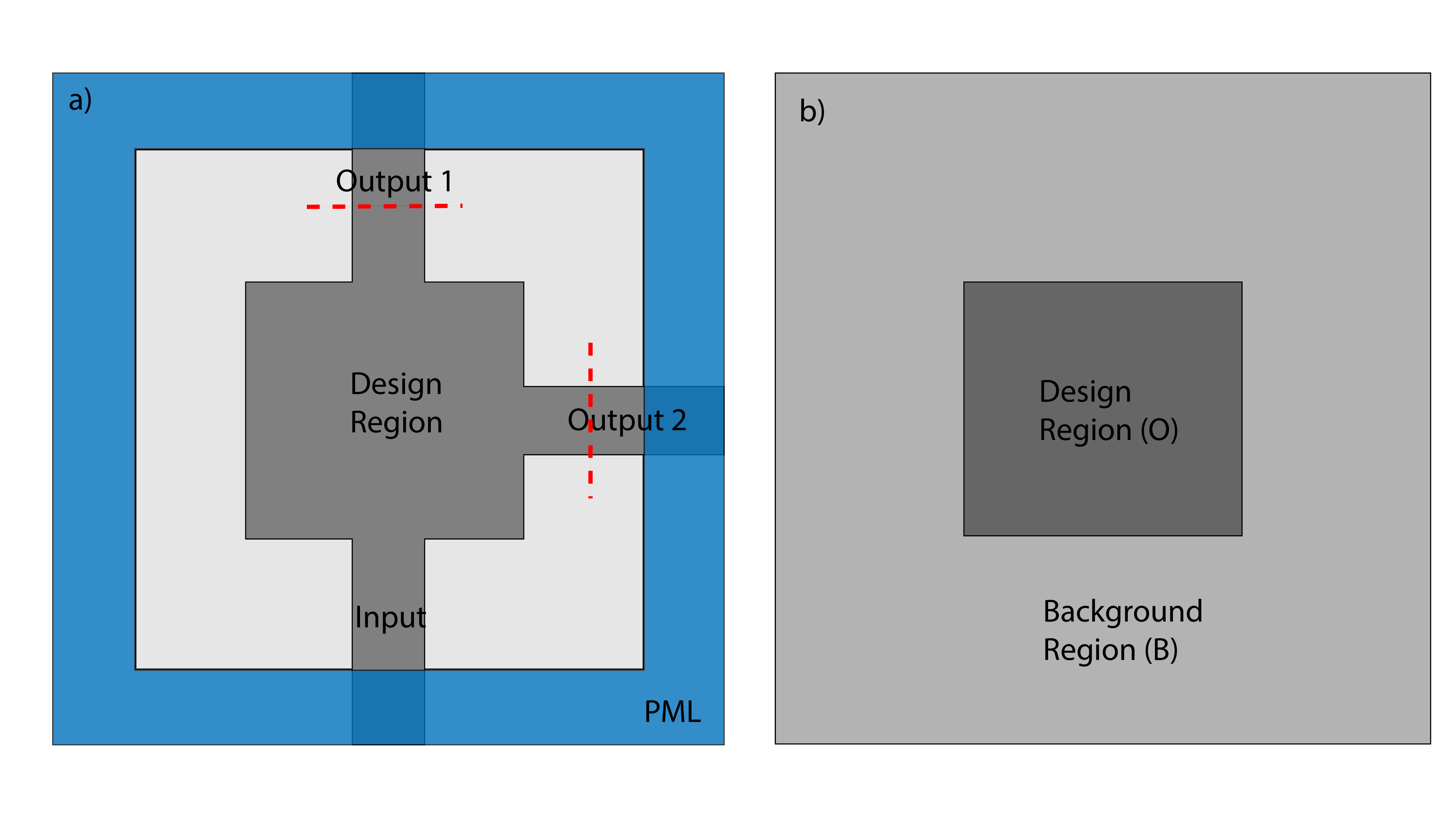}
    \caption{a) The starting structure for a waveguide power splitter. The light gray region has a relative permittivity $\epsilon_r = 2.25$, the dark gray region has a permittivity of $\epsilon_r = 6.25$. The light blue region consists of perfectly matched layer. The dotted red lines represent the lines along which we measure the objective function. b) Division of the system into the design region and the background. In the optimization process, we modify only the permittivity inside the design region. }
    \label{fig:intro}
\end{figure}

\subsection{Gradient-Based Optimization }

To optimize an optical device, we typically would like to adjust the dielectric permittivity distribution $\epsilon_r(\mathbf{r})$ to improve an objective function. In our example of Fig. 1, the objective function $L$ is a function of the two fluxes in the two output waveguides, and we will adjust the dielectric function within the design region. For this purpose we will need to compute the gradient $\partial L/\partial \epsilon_r(\mathbf{r})$. In the finite-difference setting as described above $\epsilon_r(\mathbf{r})$ is measured on a set of cells which can be integer indexed as $(p,q,s)$ in three dimensions. As a result, $\epsilon_r(\mathbf{r}) \rightarrow \epsilon_{r}(p,q,s)$. Here for notation simplicity we suppress all such spatial indices $(p,q,s)$. To proceed we note that:
\begin{equation}
    \frac{\partial L}{\partial \epsilon_r} = \frac{\partial L}{\partial \mathbf{E}}\frac{\partial \mathbf{E}}{\partial \epsilon_r}
    \label{eq:sensitivity}
\end{equation}
%+ \frac{\partial L}{\partial E^*}\frac{\partial E^*}{\partial \epsilon_r}
The gradient $\frac{\partial L}{\mathbf{\partial E}}$ can be determined analytically since the objective function $\hat L$ is typically an analytic function of $\mathbf{E}$. To determine the second term $\partial \mathbf{E}/ \partial \epsilon_r$, the derivative of a vector with respect to a scalar value, we take derivatives of Eq. \eqref{eq:forward} with respect to $\epsilon_r$ to obtain:
\begin{equation}
    \frac{\partial \mathbf{E}}{\partial \epsilon_r} = -\hat A^{-1} \frac{\partial \hat A}{\partial \epsilon_r} \mathbf{E}
    \label{eq:deriv_Ae}
\end{equation}
The left-hand side of Eq. \eqref{eq:deriv_Ae} is a vector assuming $\epsilon_r$ is isotropic  or a Jacobian matrix if it is a vector. Likewise, the derivative on the right-hand side of Eq. \eqref{eq:deriv_Ae} is a vector for an isotropic dielectric function. With a simple substitution of Eq. \eqref{eq:deriv_Ae} into Eq. \eqref{eq:sensitivity}, we get:
\begin{equation}
    \frac{\partial L}{\partial \epsilon_r} = -\bigg(\frac{\partial L}{\partial \mathbf{E}} \hat A^{-1} \bigg) \frac{\partial \hat A}{\partial \epsilon_r} \mathbf{E} = - \mathbf{E}_{adj}^T\frac{\partial \hat A}{\partial \epsilon_r}\mathbf{E}
\end{equation}
The term $\mathbf{E}_{adj}$ is the adjoint field and can be determined by solving:
\begin{equation}
    \hat A^T \mathbf{E}_{adj} = \bigg(\frac{\partial L}{\partial \mathbf{E}}\bigg)^T
    \label{eq:adjoint}
\end{equation}
where the term on the right hand side is the adjoint source. The combined solutions of $\mathbf{E}$ and $\mathbf{E}_{adj}$ are enough to compute the gradient of the objective function $\hat L$ with respect to all design variables:
\begin{equation}
    \frac{\partial L}{\partial \epsilon_r} = -\mathbf{E}_{adj}^T \frac{\partial \hat A}{\partial \epsilon_r} \mathbf{E}
    \label{eq:final_gradient}
\end{equation}
Moreover, in many numerical techniques used to solve Eq. \eqref{eq:forward} and Eq. \eqref{eq:adjoint}, such as through the use of LU decomposition, where a representation of $\hat G=\hat A^{-1}$ is acquired, one can directly reuse such a representation of $\hat  G$ to calculate the adjoint field $\mathbf{E}_{adj}$ more efficiently.
%(as opposed to reconstructing the same $\hat  G$ again).

With the gradient determined, we can then update or modify all the design variables $\epsilon_r$ in the design region accordingly:
\begin{equation}
    \epsilon_r^{(i+1)} = \epsilon_r^{(i)} +\alpha\bigg(\frac{\partial L}{\partial \epsilon_r^{(i)}}\bigg)
    \label{eq:grad_update}
\end{equation}
the superscripts $i$ and $i+1$ denote the step/epoch number in the gradient-based optimization algorithm. $\alpha$ is the learning rate hyperparameter. 

\subsection{Line Search in Gradient Descent}

In the design parameter update step of every gradient descent iteration, we see that the parameter update  in Eq. \eqref{eq:grad_update}  involves a free hyperparameter $\alpha$ called the learning rate. Line searches solve for the optimal learning rate or step size at every iteration of a gradient-based algorithm:
\begin{equation}
 \alpha^* = \min_{\alpha}(L(\epsilon_r(\mathbf{r})+\alpha\Delta\epsilon_r(\mathbf{r})))
 \label{eq:line_search}
\end{equation}
where $\Delta \epsilon_r(\mathbf{r})$ corresponds to any change in the dielectric distribution function in space, such as a gradient update as shown in Eq. \eqref{eq:grad_update}. The problem in Eq. \eqref{eq:line_search} is solved by evaluating the objective function for different $\alpha$ and taking $\alpha^*$ to be the one which generates the best improvement in the objective function. To determine $\alpha^*$ , one must sample the space of possible $\alpha$'s with sufficient density.
Moreover, for each $\alpha$, the value of $L(\epsilon_r(\mathbf{r})) + \alpha\Delta\epsilon_r(\mathbf{r})$ requires the solution or simulation of a new structure. Thus the computational cost of a line search is substantial.

In \cite{Boutami:20}, a methodology was developed which circumvented determining $\hat  G$ in Eq. \eqref{eq:solution} from scratch for each value of $\alpha$. The basis of this method is to express the solution to a line search as a perturbation theory problem. The equation that describes the system with a dielectric distribution $\epsilon_r(\mathbf{r}) + \alpha \Delta \epsilon_r(\mathbf{r})$ is:
\begin{equation}
\bigg[ \bigg(\frac{1}{\mu_0}(\nabla \times \nabla \times  )- \omega^2 \epsilon_0 \epsilon_r(\mathbf{r})\bigg) - \omega^2\epsilon_0\big(\alpha \Delta \epsilon_r(\mathbf{r})\big)\bigg]  \mathbf{E(r)} = -i\omega \mathbf{J(r)}
\label{eq:maxwell_perturb}
\end{equation}
% Consider $\alpha \Delta \epsilon_r$ as a perturbation. This modifies  Eq. \eqref{eq:maxwell} as:
By similarly discretizing this system in the frequency domain, we can solve for the new field as a function of $\alpha$: % ($\mathbf{E}(\alpha)$)
\begin{equation}
    (\hat A+\alpha \hat V)\mathbf{E}(\alpha) = -i\omega \mathbf{J}
    \label{eq:ls}
\end{equation}
where $V_{ij} = k_0^2\Delta \epsilon_r \delta_{ij}$ and we have again suppressed the spatial coordinate $\mathbf{r}$. $\delta_{ij}$ represents the Kronecker delta function. From Eq. \eqref{eq:ls}, one can derive the Lippman-Schwinger equation:
\begin{equation}
    \mathbf{E}(\alpha) = \mathbf{E}(\alpha = 0)+ \alpha(\hat G \hat V)\mathbf{E}(\alpha)
\end{equation}
where $\hat G = \hat A^{-1}$. Subsequently,  Eq. \eqref{eq:ls} can be solved using a Born series as:
\begin{equation}
	\mathbf{E}(\alpha) =\mathbf{E}(0)+ \alpha
	(\hat G\hat V)\mathbf{E}(0)+ \alpha^2 (\hat G\hat V)^2 \mathbf{E}(0)+\cdots = \sum_{k=0}^{\infty}\alpha^k(\hat G\hat V)^k \mathbf{E}(0)
	\label{eq:born_series}
\end{equation}
In Eq. \eqref{eq:born_series}, we can express the electric field for any $\alpha$  using the Green's function and the corresponding field solution for $\alpha = 0$. Therefore, if the Green's function for the current candidate structure is computed, one can use Eq.(14) to efficiently perform a line search. 

%As a result, by avoiding the full recomputation fo the Green's function, significant efficiency gains can be achieved in the context where determining $\hat  F$ is an expensive step.

% We denote a truncation to the nth term of the series in Eq: \eqref{eq:born_series} as $Q_n$. 
% follows an approximate geometric form, the use of a series accelerator called the Shanks transformation can provide a significant increase in the quality of the series estimation with only a few terms \cite{Shanks1955,Singh1991,Guerin2001}.
% The scalar version of the Shanks transform has the form:

In Ref. \cite{Boutami:20}, it was noted that the convergence of the series in Eq. \eqref{eq:born_series} is typically poor. On the other hand, it was also noted that the sum in the Eq. \eqref{eq:born_series} can be efficiently estimated using the technique of the Shank transformation. We denote the sum of the first $n$ terms in Eq. \eqref{eq:born_series} as $\mathbf{E}_n(\alpha)$. The Shank's transformation \cite{Shanks1955,Singh1991,Guerin2001} of the series $\mathbf{E}_n(\alpha)$ has the form: 

\begin{equation}
    \mathbf{T}(\mathbf{E}_n(\alpha)) = \frac{\mathbf{E}_{n+2}(\alpha)\mathbf{E}_n(\alpha) -\mathbf{E}_{n+1}(\alpha)^2}{\mathbf{E}_{n+2}(\alpha)-2\mathbf{E}_{n+1}(\alpha)+\mathbf{E}_n(\alpha)}
    \label{eq:shanks}
\end{equation}
where $\mathbf{T}(\mathbf{E}_n)$ is an element-wise transformation of the vector $\mathbf{E}_n$, and all operations in Eq. \eqref{eq:shanks} are performed element-wise. $\mathbf{T}(\mathbf{E}_n)$ provides a very accurate estimation of the infinite sum in Eq. \eqref{eq:born_series}. Therefore, the use of Shank's transformation can allow us to perform a very efficient line search.

\subsection{Reducing the Cost of a Line Search via a  Schur Complement}

With the use of the Shank's transformation technique, the key computational cost in the line search becomes the evaluation of the Green's function in each term of Eq. \eqref{eq:born_series} for each candidate structure. To reduce the cost associated with evaluating Eq. \eqref{eq:born_series}, we observe that in typical inverse design problems, the design region consists of a limited subset of all degrees of freedom being simulated. This observation can be exploited to reduce the computational cost of determining the Green's function for each candidate structure \cite{Zhao:19}. To do this, we partition the degrees of freedom on the grid into two subdomains. One consists of the design region and $\Omega_i$, and the other consists of a background which is the region of the computational domain outside of the design region. (See Fig. 2 for an illustration). Eq. \eqref{eq:forward} can then be written in the form as: 

\begin{equation}
    \begin{bmatrix}
   \hat A_O  & \hat A_{OB} \\
   \hat A_{BO} &  \hat A_B
   \end{bmatrix}\begin{bmatrix}
   \mathbf{E}_O \\
   \mathbf{E}_B
   \end{bmatrix} = -i\omega \begin{bmatrix}
   \mathbf{J}_O \\
   \mathbf{J}_B
   \end{bmatrix}
   \label{eq:partitioned_A}
\end{equation}
The submatrices $\hat A_O$ and $\hat A_B$ are now the part of the matrix $\hat A$ restricted to the design region and the background, respectively. The off-diagonal blocks $\hat A_{OB}$ and $\hat A_{BO}$ represent the couplings between grid points of these two regions. This coupling occurs between the design region and the background region. $\mathbf{J}_B$ and $\mathbf{J}_O$ represent the sources in either subdomain. 

Using the second row of Eq. \eqref{eq:partitioned_A}, we obtain:
\begin{equation}
     \hat A_{BO}\mathbf{E}_O + \hat A_B\mathbf{E}_B = -i\omega \mathbf{J}_B
     \label{eq:second_row}
\end{equation}
We can solve for $\mathbf{E}_B$ in Eq. \eqref{eq:second_row} and substitute the solution back into the first row of Eq. \eqref{eq:partitioned_A} to get an equation for $\mathbf{E}_O$ only:
\begin{equation}
    \hat S\mathbf{\mathbf{E}_O} = -i\omega \mathbf{J}_S%\big(\mathbf{J}_O - A_{OB}A_B^{-1} \mathbf{J}_B\big),
    \label{eq:reduced}
\end{equation}
where:
\begin{equation}
\hat S \equiv \hat A_O - \hat A_{OB}\hat A_B^{-1}\hat A_{BO}
\label{eq:schur}
\end{equation}
\begin{equation}
    \mathbf{J}_S \equiv \mathbf{J}_O - \hat A_{OB} \hat A_B^{-1}\mathbf{J}_B. 
    \label{eq:schur_source}
\end{equation}
Below, we will refer to Eq. \eqref{eq:partitioned_A} as the ``original'' system and Eq. \eqref{eq:schur} as its corresponding ``reduced'' system. Because $\hat S$ will typically be much smaller in dimension than $\hat A$, we can expect solving the linear system with $\hat S$ to be much faster in many situations as discussed in Ref. \cite{Zhao:19}

We now show that one can perform the line search using the reduced system as opposed to the original system. Since the only modification of the structure occurs in the design region, within each line search we need to solve for

\begin{equation}
    (\hat S+\alpha \hat V_S)\mathbf{E}_O(\alpha) = -i\omega \mathbf{J}_S
    \label{eq:ls_schur}
\end{equation}
where $V_S$ is the same as $V$ but is reduced so the dimension matches that of $\hat S$.
The solution of this equation can be written analogous to Eq. \eqref{eq:born_series}) as:

\begin{equation}
	\mathbf{E}_O(\alpha) =\mathbf{E}_O(0)+ \alpha
	(\hat G_S\hat V_S)\mathbf{E}_O(0)+ \alpha^2 (\hat G_S\hat V_S)^2 \mathbf{E}_O(0)+\cdots = \sum_{k=0}^{\infty}\alpha^k(\hat G_S\hat V_S)^k \mathbf{E}_O(0)
	\label{eq:schur_born_series}
\end{equation}
where $\hat G_S = \hat S^{-1}$. Thus, the technique of Shank's transformation can be applied to the Born series as well. Therefore, we have shown that the techniques of Schur's complement and Shank transformation can be combined to further reduce the computational cost associated with the line search. In the next section, we will provide a numerical demonstration of such a combination for an optimization problem in photonics.

\section{Results}
We consider a two-dimensional problem of a mode splitter introduced in Section 1. 
In the center is a design region with a size of $2 \times 2$ $\mu$m. We simulate this device
for the electric field polarization perpendicular to the 2D plane ($E_z$, $H_x$, $H_y$), at wavelength $\lambda = 1.55$ $\mu$m. Further specifications can be found in Table 1.

%Table \ref{Tab:specs} (Table \ref{Tab:times}, Table \ref{tab:fillin})

\begin{figure}
    \centering
    \includegraphics[width = 5 in]{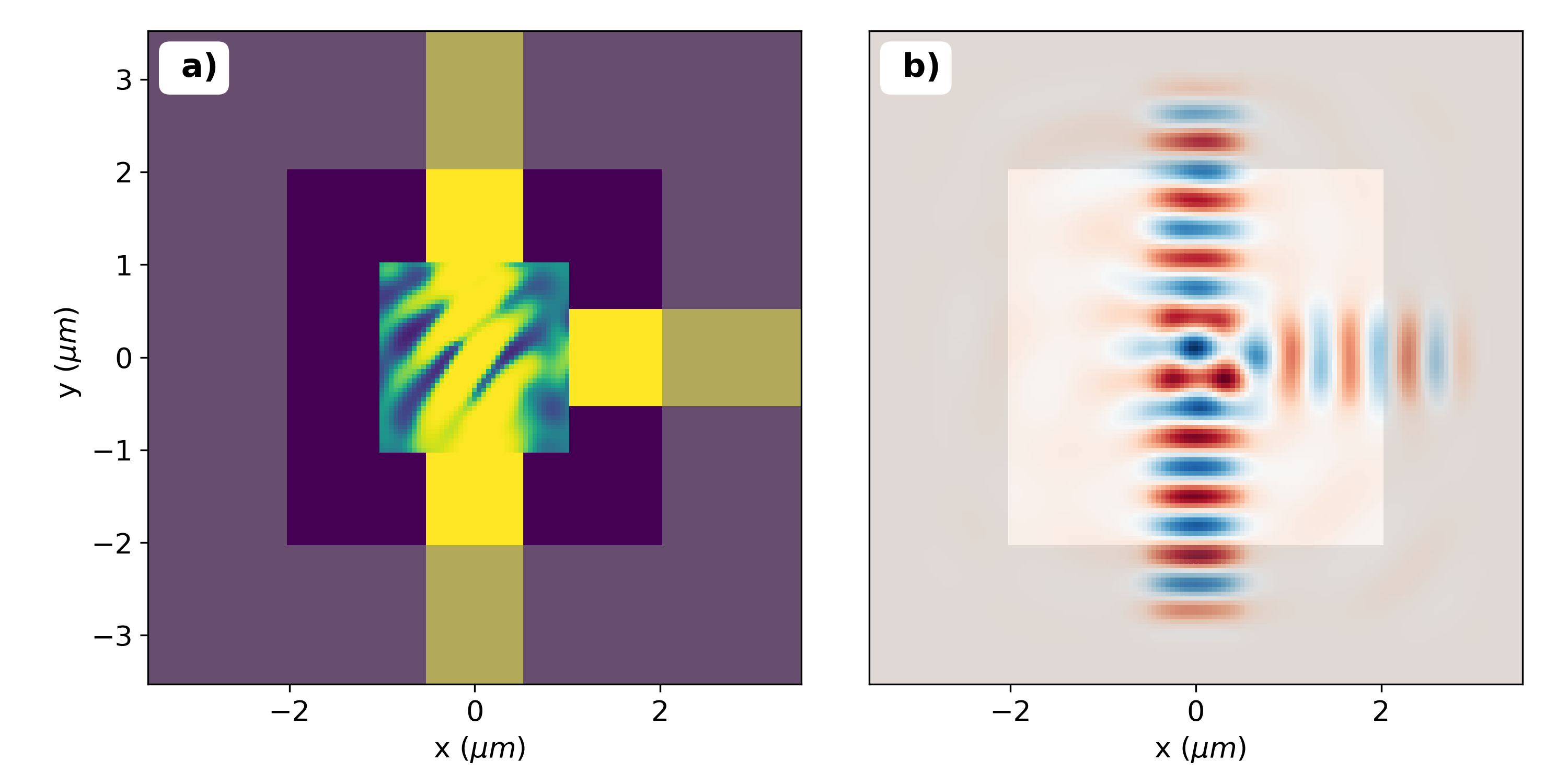}
    \caption{a) The spatial distribution of the dielectric $\epsilon_r$ of an optimized mode power splitter. b) The corresponding real profile of the $E_z$ field as a function of $x$ and $y$ for the fully optimized device.}
    \label{fig:splitter_figure}
\end{figure}

\begin{table}[H]
\begin{center}
\begin{tabular}{ |c|c| } 
 \hline
  Descriptor & Values \\ \hline 
 $(N_x,N_y)$ &  (141,141) \\ 
 $x_{range}$ & (-3.525  3.525) $\mu$m \\ 
 $y_{range}$ & (-3.525  3.525) $\mu$m \\ 
 $dx = dy$ & 0.05 $\mu$m \\
 design region & (-1.0, 1.0) $\mu$m \\
 PML thickness & (1.5, 1.5) $\mu$m \\
 \hline
\end{tabular}
\vspace{0.1 in}
\caption{Test system specifications. $N_x$,$N_y$ represent the number of grid cells along the $x$ and $y$ dimension respectively, $x_{range}$, and $y_{range}$ represent the physical domain length corresponding to the grid. $dx$, $dy$ is the discretization size for each grid cell. The design region is a square, and in the table we provide the minimum and maximum $x$ or $y$ coordinates of the design region. In the row labelled PML thickness, we provide the width of the PML layer in the $x$ and $y$ direction.}
\end{center}
\label{Tab:specs}
\end{table}

% The objective function $L(\epsilon_r)$ for this mode converter is proportional to the overlap integral involving the electric field along a line perpendicular to the output waveguide, and the spatial profile of the TE$_1$ magnetic field mode of the output waveguide:
%In this problem 8.5\% of all simulated degrees of freedom are being optimized (1681/19881). 

The objective function for this power splitter is related to the overlap integral of the simulated field with the $\text{TE}_0$ modes propagating in the two output waveguide arms. For each arm normal to the $x$ and $y$ direction, as labelled by $i=x,y$:
\begin{equation}
   L_i(\epsilon_r) = A_i\bigg|\int_{\Omega_{i}}{E_{z}H_{i,\text{TE}_0}^*(x,y)+ E_{z,\text{TE}_0}^*(x, y)H_{i}(x,y) di \bigg|^2}
    \label{eq:FoMwaveguide_arm}
\end{equation}
where $H_{i,\text{TE}_0}(x,y)$ is the $H_x$ or $H_y$ field of the $\text{TE}_0$ mode of the waveguide $i$ (and likewise for $E_{z,\text{TE}_0}(x,y)$. An asterisk denotes taking the complex conjugate and $E_z$ and $H_i$ correspond to the field obtained by the simulation using the current iteration of the device. $\Omega_i$ denotes an integration line that is perpendicular to waveguide $i$. $A_i$ is a normalization constant as defined by:
\begin{equation}
    A_i = 1/(8\int_{\Omega_i}\text{Re}\big(E_{z, TE_0}(x,y)H_{i,TE_0}^*(x,y)\big) di)
\end{equation}
Using the $L_i$'s, we define an objective function $L(\epsilon_r)$ as:
\begin{equation}
   L(\epsilon_r) = 4(L_1(\epsilon_r)L_2(\epsilon_r))
    \label{eq:FoMwaveguide}
\end{equation}
$L(\epsilon_r)$ reaches a maximum value of $1$ when the power is evenly split between the two arms with zero loss. The derivative of $L_i$ with respect to the dielectric permittivity $\epsilon_r$:
\begin{equation}
    \frac{\partial L_i}{\partial \epsilon_r} = 2 k_0^2 \text{Re}\bigg((E_z(x,y) {E}_{z, adj}(x,y)) L_i^*(\epsilon_r)\bigg)
\end{equation}
where ${E}_{z, adj}$ is the $z$-component of the adjoint field as derived in Eq. \eqref{eq:adjoint}. $k_0 = 2\pi/\lambda$. 

In order to account for the fact that $\epsilon_r$ is bounded above and below in typical optimization problems, we modify the gradient:
\begin{equation}
    \text{proj}\bigg(\frac{\partial L}{\partial \epsilon_r}\bigg) = \frac{\partial L}{\partial \epsilon_r}\begin{cases}
        \frac{(\epsilon_r - \epsilon_{min})}{\max(|\frac{\partial L}{\partial \epsilon_r}|)}, & \frac{\partial L}{\partial \epsilon_r} \geq 0 \\
        \frac{(\epsilon_{max} - \epsilon_{r})}{\max(|\frac{\partial L}{\partial \epsilon_r}|)}, & \frac{\partial L}{\partial \epsilon_r} < 0 \\
    \end{cases}
    \label{eq:search_direction}
\end{equation}
In our design, we choose $\epsilon_{min}= 2.25$ and $\epsilon_{max} = 6.25$. Eq. \eqref{eq:search_direction} amounts to a projection of the gradient based on the lower and upper bounds on the design parameter values \cite{Drummond2004}. The projection of the gradient allows us to effectively reparametrize the space of the hyperparameter $\alpha$ to be between 0 and 1. As a result, the line search in Eq. \eqref{eq:line_search} becomes a constrained optimization problem based on the bound constraints of the design parameters.

For the Shanks transforms as described in Eq. \eqref{eq:shanks}, we use $n=3$.
\begin{figure}[H]
    \centering
    \includegraphics[width = 3.5 in]{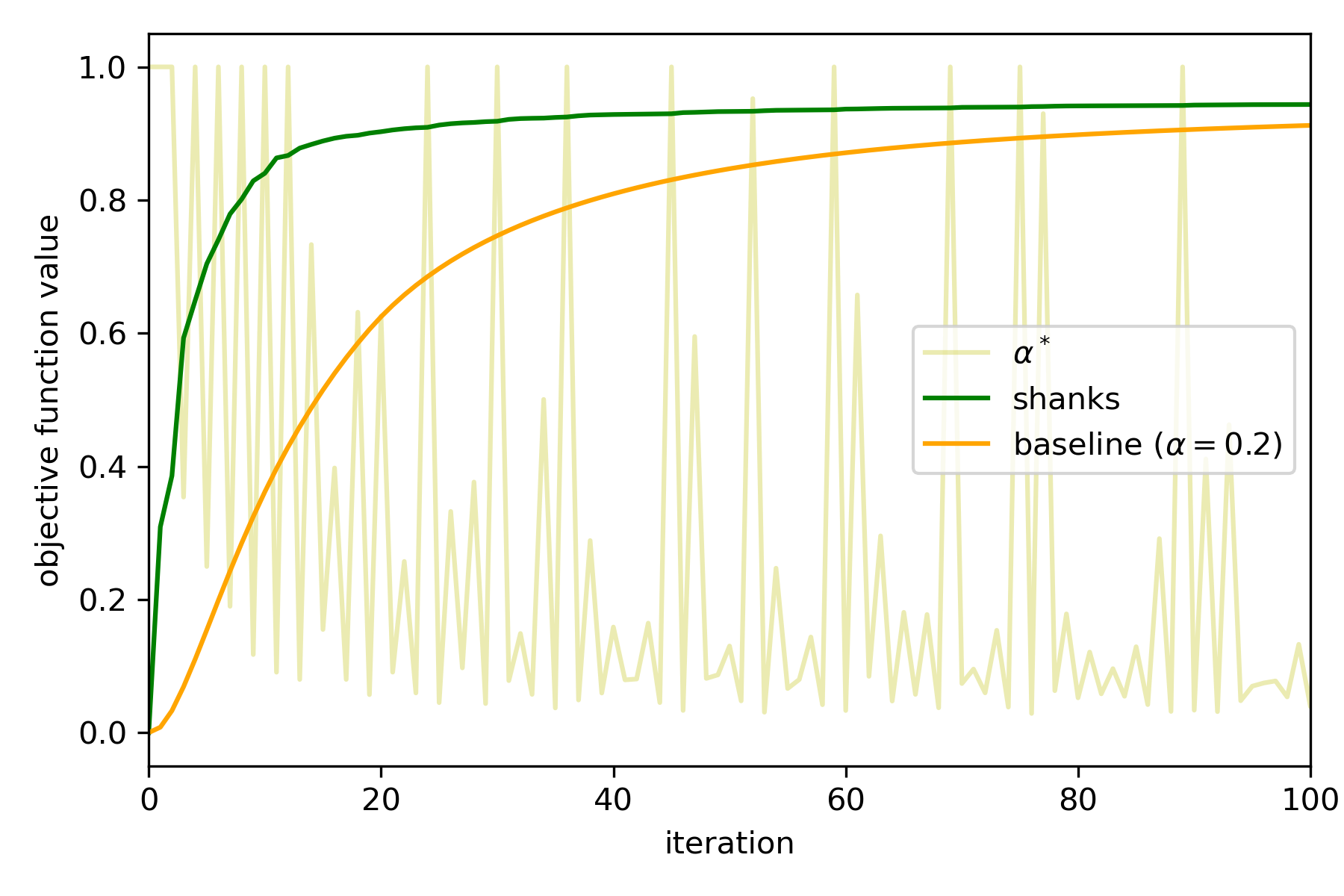}
    \caption{The green line represents the value of the objective function versus the number of iterations, where each iteration the optimal hyperparameter $\alpha^*$ is used. The light gold line represents the optimal $\alpha^*$ determined at each iteration. The orange line is essentially the same optimization as the green line, except that a constant $\alpha = 0.2$ is used for each iteration.
    \eqref{eq:search_direction}}
    \label{fig:iteration_history}
\end{figure}
In Fig. \ref{fig:iteration_history}, we compare the objective function with respect to the number of iterations for different learning rate strategies. The orange line is our baseline, which is a simple gradient descent with a constant fixed $\alpha = 0.2$, chosen so that the objective increases monotonically throughout the optimization process. The green line represents the same simple gradient descent but the learning rate is optimized using of the born series accelerated line search. The light gold line represents the optimal learning rate determined at each iteration.  Clearly, the use of the line search accelerates the convergence with respect to the number of iterations. Also, we see that the learning rate can vary significantly throughout the optimization through the entire possible range of 0 to 1. However, Fig. \ref{fig:iteration_history} does not compare the amount of time it takes for the optimization to complete.

\begin{figure}[H]
    \centering
    \includegraphics[width = 4 in]{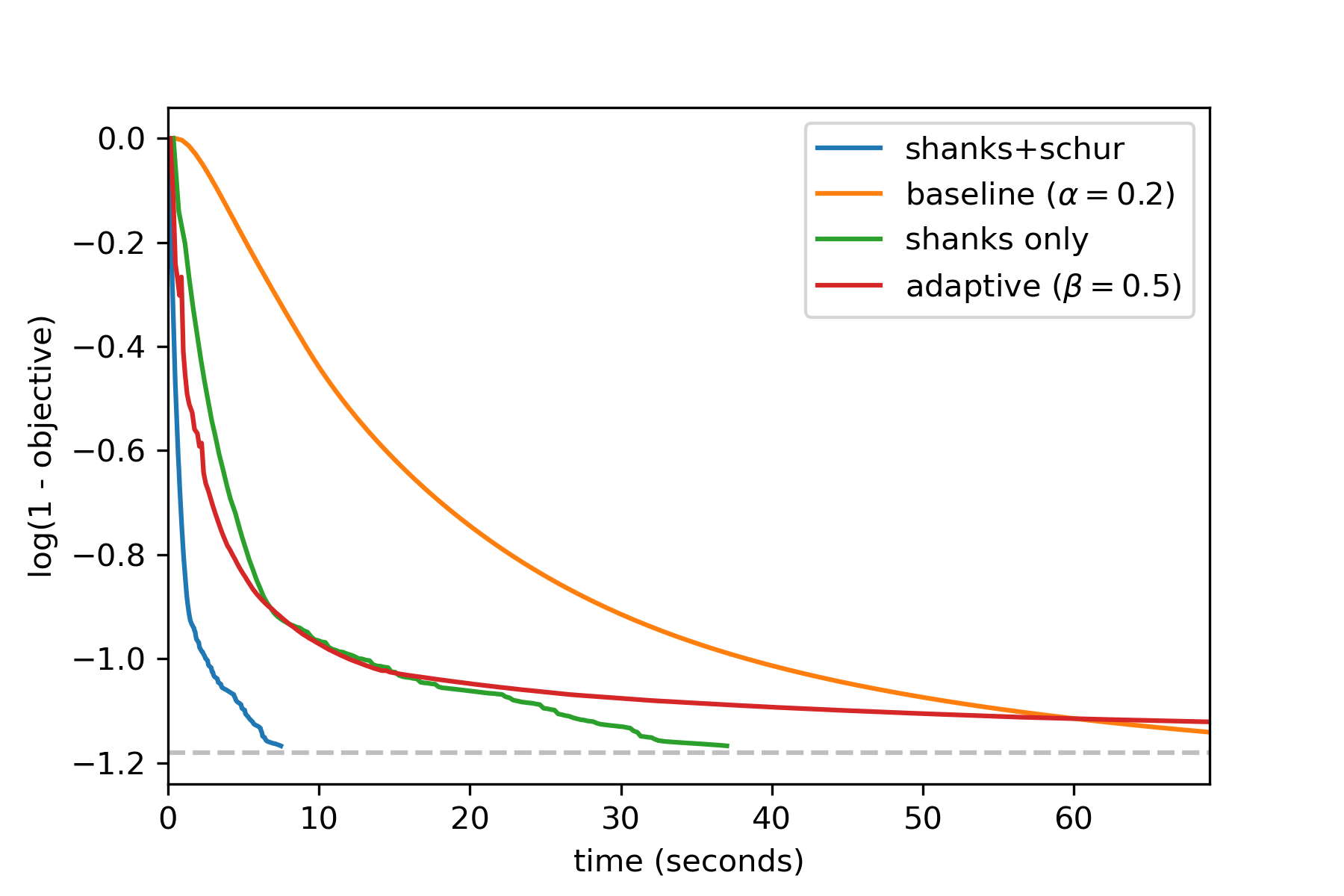}
    \caption{Comparison of the time taken by a line search enhanced gradient descent (blue), gradient descent with our accelerated line search (orange) and then two baselines using gradient descent with a constant learning rate (green) and one with an adaptively decayed learning rate (red).}
    \label{fig:Timing}
\end{figure}

In Fig. \ref{fig:Timing}, we compare the full time cost taken by our method compared to a variety of baselines. In this way, we can now evaluate the performance of our combined line search and Schur complement method.
The green and orange lines correspond to the same setups as in Fig. \ref{fig:iteration_history}. The red line is an "adaptive" gradient descent method \cite{ge2019step,you2019does} whereby we start out with $\alpha=1$ but successively decrease it by a geometric factor $\beta$ whenever an update of the parameters causes a decrease in the objective function. Note that there is still a hyperparameter here in choosing the size of the geometric decay factor $\beta$. In comparing such an adaptive method (red line), with the exact line search (orange line), we note that while the initial performance of the adaptive method is good, in the later times the adaptive method underperforms as compared to the exact line search. 
This is because in the adaptive method the learning rate decreases as a function of time, but as we can see in Figure \ref{fig:Timing}, the optimal learning rate may still be large in later iterations. The performance of the exact line search based on the Shank transformation can be further improved by combining it with the Schur domain decomposition that reduces the size of the computational domain at each iterations, as can be seen by comparing the green line with the blue line.

\begin{table}
\begin{center}
\begin{tabular}{ |c|c|c| } 
 \hline
  & Original System ($\hat A$) & Reduced system ($\hat S$) \\ \hline
 LU factorization & 0.11 s & 0.01 s \\ 
 Back-substitution & 0.01 s & 0.001 s \\ 
  Line search & 0.03 s& 0.003 s \\ 
 \hline
\end{tabular}
\vspace{0.1 in}
\caption{Measured times for different processes (measured in seconds (s)) within a single iteration step when performing the method of exact line search. The LU factorization corresponds to factorizing $\hat A$ in Eq. \eqref{eq:forward} and $\hat S$ in Eq. \eqref{eq:reduced}.  Calculations were done in MatLab2019 on a 2.4 GHz Intel Core i9 CPU}
\label{Tab:times}
\end{center}
\end{table}

In Table 2, we provide a more detailed examination of the time cost of various steps required when an exact line search is performed. The original system corresponds to the original system matrix A as described in Eq. (2). The reduced system corresponds to the case where the Schur domain decomposition method has been applied to generate a reduced system matrix $\hat S$, as described in Eq. (18). For each iteration step when performing the exact line search, the major components are the “LU decomposition” of the system matrix, the “back substitution” after the LU decomposition has been performed, and the application of the Shank transformation which determines the optimal $\alpha$ in the “line search”. (The quotations here correspond to various entries in Table 2).  We note that performing domain decomposition significantly reduces the time cost for every major component in an iteration step. 
 
In comparing the usual gradient descent methods, with the exact line search, we note that while the exact line search reduces the number of iterations required as shown in Figure 3, the computational cost for each iteration step is higher. Thus, in applying the method of exact line search, there is a particular advantage of using the domain decomposition method. We also note that in applying the domain decomposition method there is an initial overhead in forming the reduced system matrix $\hat S$. In our case, the cost of such overhead is not significant since the number of iteration steps is sufficiently large.

To further understand how the use of domain decomposition speeds up the line search algorithm,  we can look at the numerical properties of the matrices that come out of Eq. \eqref{eq:forward} and Eq. \eqref{eq:schur}.  Numerically, these equations are solved through LU decomposition and back-substitution. Because of the fact that the matrices are sparse, Eq. \eqref{eq:forward} and Eq. \eqref{eq:reduced} are solved using sparse direct solvers \cite{Davis2004,Davis2006,Davis2016} with an optimized permutation algorithm to minimize fill-in in the factorization. Furthermore, we use a perfect electric conductor/perfect magnetic conductor (PEC/PMC) boundary (rather a periodic boundary) condition on the boundary edges that are covered with a perfectly matched layer (PML) \cite{Berenger1994}.

Now, assume we have a factorization $\hat L$, $\hat U$ for $\hat A$ and $\hat L_S$, $\hat U_S$ for $\hat S$. To evaluate the numerical cost of forming Eq.\eqref{eq:born_series} or Eq. \eqref{eq:schur_born_series}, as is required for performing the exact line search, we need to determine the cost of back-substitution in the form of $ U^{-1}(L^{-1}b)$ where $b$ is a vector. Thus, the time to determine the Born series in Eqs. \eqref{eq:born_series} and \eqref{eq:schur_born_series} is proportional to the non-zeros in the LU factors of $\hat A$ and $\hat S$. In general, we expect $\hat S$ to have less fill-in than $\hat A$ for two reasons. One, $\hat S$ is smaller in dimension than $\hat A$, and two, $\hat S$ is block sparse with fewer non-zeros than $\hat A$, which further reduces the degree of fill-in during the factorization process. For the system studied in this paper, the reduced system itself actually contains a fewer number of nonzero elements than the original system as illustrated in Table \ref{tab:fillin}. Thus, for the reduced system, the LU factors of the system matrix have far less non-zero elements as compared with that of that the original system. It is for this reason that the time cost of each iteration step of exact line search is far lower in the reduced system.

\begin{table}[H]
\begin{center}
\begin{tabular}{ |c|c|c| } 
 \hline
  & Original System ($\hat A$) & Reduced System ($\hat S$) \\ \hline 
 linear system matrix & 96609 & 33361 \\ 
 L/U factor of the linear system matrix & 434960 & 48610 \\ 
 \hline
\end{tabular}
\vspace{0.1 in}
\caption{Number of nonzeros in the system matrix and LU factors of the original and reduced systems for the system described in Tab. \ref{Tab:specs}.}
\end{center}
\label{tab:fillin}
\end{table}

\section{Conclusion}
In this paper, we have demonstrated that the method of exact line search, as enabled by applying the Shank transformation on a Born series resulting from the Lippman Schwinger formalism, can be further speed up with the use of a Schur domain decomposition method. For future works, we note that our technique can be further enhanced by minimizing the evaluations needed to find the optimal learning rate, such as by using a back-tracking line search \cite{boyd2004}. Additionally, this technique has significant implications for quasi-newton methods such as limited memory Broyden–Fletcher–Goldfarb–Shanno algorithm (L-BFGS) \cite{Nocedal1980}, where line searches take up the majority of the time in each iteration. Because quasi-newton methods require a line search to be done every iteration, our technique can effectively render every iteration of a L-BFGS the same cost as normal first order methods.

This work is supported by MURI projects from the U. S. Air Force Office of Scientific Research (AFOSR) (Grant No. FA9550-17-1-0002 and FA9550-21-1-0244).

\bibliographystyle{unsrt}  
\bibliography{references}  %%% Remove comment to use the external .bib file (using bibtex).

\end{document}